# On-Chip Single Plasmon Detection

Reinier W. Heeres[1,*], Sander N. Dorenbos[1], Benny Koene[1], Glenn S. Solomon[2],

Leo P. Kouwenhoven[1], Valery Zwiller[1]

[1]Kavli Institute of Nanoscience, Delft University of Technology,
P.O. Box 5046, 2600 GA Delft, The Netherlands

[2]Joint Quantum Institute, NIST and University of Maryland,
100 Bureau Drive MS-8423, Gaithersburg, MD 20899-8423 USA

[*]r.w.heeres@tudelft.nl

**ABSTRACT**

**Surface plasmon polaritons (plasmons) have the potential to interface electronic and optical devices. They could prove extremely useful for integrated quantum information processing. Here we demonstrate on-chip electrical detection of single plasmons propagating along gold waveguides. The plasmons are excited using the single-photon emission of an optically emitting quantum dot. After propagating for several micrometers, the plasmons are coupled to a superconducting detector in the near-field. Correlation measurements prove that single plasmons are being detected.**

The term *plasmons* refers to light confined to a metal/dielectric interface, with appealing characteristics of shortened wavelengths and enhanced field strengths[1]. Plasmons are also easily guided on-chip over distances of many micrometers. Most interest has been in the classical, many-photon regime where plasmons are generated by light and after traveling along the metal surface, re-emitted as photons into free space. A few experiments with respect to quantum properties have been performed. First, it was shown that the plasmon-mediated enhanced light transmission through arrays of holes in a metal surface conserves the photon's polarization properties, including quantum superpositions[2]. Similar work showed that energy-time entanglement is also preserved[3]. Second, it was shown that coupling single emitters to silver nanowires in the near-field allows excitation of single plasmons[4,5]. All these schemes relied on conversion of the plasmon to a free photon and subsequent far-field detection with traditional single photon detectors. Alternatively, on-chip electrical detection has been demonstrated using organic photodiodes[6], Gallium Arsenide structures[7] and Germanium wires[8]. However, none of these techniques has provided single plasmon sensitivity. By coupling a plasmon waveguide to a Superconducting Single-Photon Detector (SSPD), we demonstrate on-chip electrical detection of single plasmons. Next to single-photon sensitivity[9], this on-chip, near-field detection has the potential for high detection efficiency, high band-width and low timing jitter.

Our SSPDs consist of a meandering NbN wire (~100 μm long, 100 nm wide, ~5 nm thin). The critical temperature $T_c$ below which the wire becomes superconducting is approximately 9 K. Absorption of a single photon is sufficient to create a local region in the normal, resistive state. This short-lived resistive state is detected as a voltage pulse at the terminals of the wire. The excess energy is dissipated within a fraction of a nanosecond, after which the superconducting state can be restored. The detection rate is limited by the kinetic inductance of the superconducting wire[10], which in our current detectors gives a maximum count-rate of 100 MHz. As we show here, this detection mechanism can also efficiently measure individual plasmons.

We fabricate plasmon waveguides from polycrystalline gold strips, which are electrically insulated from the NbN by a thin dielectric (Fig. 1a, Supporting Information). Gratings at both ends serve to couple incoming free-space photons to plasmons confined to the bottom gold/dielectric interface[11]. These plasmons propagate to the detector where they are absorbed and detected (Fig. 1b).

Measurements are performed in a cryostat at ~4 K with the sample mounted on an XYZ

translation stage. A laser is focused through a cold microscope objective and a count-rate XY-map is measured as a function of laser-spot position. We find a large detector response when the laser directly illuminates the SSPD (3 peaks at X ≈ 11 µm in Fig. 1c and 1d). The detector response is very low with the laser spot on the substrate or gold strip, except at the grating regions. These detector peaks (consistently shifted for waveguides 1 and 2, Fig. 1c) show that light converted to plasmons is detected electrically on-chip.

To substantiate the electrical detection of plasmons we have performed several checks. First, we measured one waveguide with an intentional 1 µm gap between grating and detector (Supporting Fig. S1), resulting in a strong suppression of the detector signal. Second, we rotated the incoming light polarization and retrieve a proper polarization dependent detector signal (Supporting Fig. S2). We further measured wavelength dependence and find a vanishing signal at ~650 nm due to losses in the gold film, and a rapidly increasing detector response for longer wavelengths. Finally we measured the plasmon decay length in our gold strips and find a 1/e-decay of ~ 10 µm for 810 nm light. All these checks are in qualitative agreement with simulations using Lumerical FDTD software.

We have fabricated several waveguides with varying grating-detector distance and shapes (e.g. a bend, Supporting Fig. S1). We also designed more complex structures to illustrate the flexibility of our fabrication method, one of which is a Y-splitter (Fig. 2a). Again the experiment consists of scanning a laser beam across the sample. In this case, the count-rates of the left (Fig. 2b) and the right (Fig. 2c) detectors are monitored simultaneously. Next to the individual detectors, which are visible in just one of the signals, the grating in the bottom left is visible in both images. This indicates that plasmons excited at the grating are propagating in both arms of the Y structure. This device could be used as an integrated plasmon Hanbury-Brown-Twiss interferometer. The splitter is designed to be symmetric and therefore balanced. However, because the efficiency of the individual detectors is determined by their intrinsic sensitivity (e.g. microscopic details) and the applied bias current, it is not possible to measure the absolute splitting ratio in this configuration. Our current plasmon waveguides are all multi-mode, but in the future they can be downsized to single-mode structures implementing interference-based devices such as Mach-Zehnder interferometers[12] and coincidence-based quantum logic gates[13] using plasmons.

SSPDs are well characterized and proven to have single photon sensitivity[9]. The power dependence (Supporting Fig. S3) already strongly suggests single plasmon sensitivity, since it is linear even in the regime where the average number of incoming photons within the detector dead-time (~10 ns) is much smaller than one. However, to unambiguously prove single plasmon sensitivity requires a single photon source for plasmon excitation in addition to a time-correlation measurement[4]. We performed this measurement by cooling Stranski-Krastanov (SK) quantum dots (QDs) in a second cryostat. This sample contains QDs at the center of a distributed Bragg reflector microcavity (Fig. 3b, inset) to enhance the brightness[14]. The emission of one of these QDs is collected and sent through a narrow bandpass filter. The filtered spectrum is shown in Fig. 3b. Ten percent of this emission is collected in an optical fiber and sent to an Avalanche Photo Diode (APD). The other 90 percent is coupled through free-space to the setup containing the SSPD with waveguides (Fig. 3a). To confirm that we are looking at a single-photon source, we first position the incoming beam directly at the SSPD. The correlation measurement between detection events from the APD and the SSPD is shown in figure 3c. An anti-bunching dip with a fitted depth close to 0.5 is clearly visible. The fact that the dip does not go below the theoretical limit of 0.5 for a single emitter is likely to be due to background emission from the substrate and non-perfect filtering, especially since the two main peaks in the spectrum only constitute ~60% of the total counts.

The next step is to perform another XY-scan using the single-photons from the QD (Fig. 3d). This scan clearly confirms that single plasmons can be excited by illuminating the gratings with single photons. After focusing the emission of the single-photon source on the lower-right grating another correlation measurement is performed between the APD and the SSPD. The SSPD now detects plasmons that have propagated about 7.5 µm along the waveguide after coupling in through the grating. The resulting data clearly shows that the quantum statistics of the original photons are maintained and that single plasmons are being detected. This is the first time that single plasmons are observed on-chip in the near-field, and opens up a wide range of possibilities. By placing a single emitter on-chip using nanomanipulation[15], the integration could even be taken one step further, resulting in a complete optical circuit with efficient coupling of a single-photon source[16,17] to a waveguide and a detector on a monolithic device.

Detecting single plasmons on-chip makes our waveguide-detector scheme very promising for ultra-fast detection with low dark-count rates. The time-resolution of < 80 ps that can be achieved with an SSPD today is comparable to the best APDs. However, contrary to those detectors the SSPDs also provide good sensitivity in the near infra-red range, up to several micrometers in wavelength. Combined with the flexibility for fabricating various complex waveguide structures this results in many potential applications as sensors or interconnects and for quantum information processing.

**Acknowledgements** We thank Ewold Verhagen and Freek Kelkensberg for discussions and Nika Akopian for support. This work is supported financially by The Netherlands Organisation for Scientific Research (NWO/FOM).

**Supporting Information** Materials and Methods (fabrication, measurement setup, quantum dots, fitting details). Additional device geometries and measurements (2D scan, power dependence, polarization dependence).

**References**

1. Barnes, W.L.; Dereux, A.; Ebbesen, T.W. *Nature* **2003,** 424, 824–830

2. Altewischer, E.; Van Exter, M.P.; Woerdman, J.P. *Nature* **2002**, 418, 304–306

3. Fasel, S.; Robin, F.; Moreno, E.; Erni, D.; Gisin, N.; Zbinden, H. *Phys. Rev. Lett.* **2005**, 94, 110501

4. Akimov, A.V.; Mukherjee, A.; Yu, C.L.; Chang, D.E.; Zibrov, A.S.; Hemmer, P.R.; Park, H.; Lukin, M.D. *Nature* **2007**, 450, 402–406

5. Kolesov, R.; Grotz, B.; Balasubramanian, G.; Stöhr, R.J.; Nicolet, A.A.L.; Hemmer, P.R.; Jelezko, F.; Wrachtrup, J. *Nature Physics* **2009**, 5, 470 - 474

6. Ditlbacher, H.; Aussenegg, F.R.; Krenn, J.R.; Lamprecht, B.; Jakopic, G.; Leising, G. *Appl. Phys. Lett.* **2006**, 89, 161101

7. Neutens, P.; Van Dorpe, P.; De Vlaminck, I.; Lagae, L.; Borghs, G. *Nature Photonics* **2009**, 3, 283–286

8. Falk, A.L.; Koppens, F.H.L.; Yu, C.L.; Kang, K.; de Leon Snapp, N.; Akimov, A.V.; Jo, M.H.; Lukin, M.D.; Park, H. *Nature Physics* **2009**, 5, 475–479

9. Gol'tsman, G.N.; Okunev, O.; Chulkova, G.; Lipatov, A.; Semenov, A.; Smirnov, K.; Voronov, B.; Dzardanov, A.; Williams, C.; Sobolewski, R. *Appl. Phys. Lett*. **2001**, 79, 705

10. Kerman, A.J.; Dauler, E.A.; Keicher, W.E.; Yang, J.K.W.; Berggren, K.K.; Gol'tsman, G.N.; Voronov, B. *Appl. Phys. Lett.* **2006**, 88, 111116

11. Verhagen, E.; Polman, A.; Kuipers, L. *Opt. Express* **2008**, 16, 45–57

12. Bozhevolnyi, S.I.; Volkov, V.S.; Devaux, E.; Laluet, J.Y.; Ebbesen, T.W. *Nature* **2006**, 440, 508–511

13. Politi, A.; Cryan, M.J.; Rarity, J.G.; Yu, S.; O'Brien, J.L. *Science* **2008**, 320, 646–649

14. Solomon, G. S.; Pelton, M.; Yamamoto, Y. *Phys. Rev. Lett.* **2001**, 86, 3903–3906

15. van der Sar, T.; Heeres, E.C.; Dmochowski, G.M.; de Lange, G.; Robledo, L.; Oosterkamp, T. H.; Hanson, R. *Appl. Phys. Lett.* **2009**, 94, 173104

16. Chang, D.E.; Sørensen, A.S.; Hemmer, P.R.; Lukin, M.D. *Phys. Rev. Lett.* **2006**, 97, 053002

17. Chang, D.E.; Sørensen, A.S.; Demler, E.A.; Lukin, M.D. *Nature Physics* **2007**, 3, 807–812

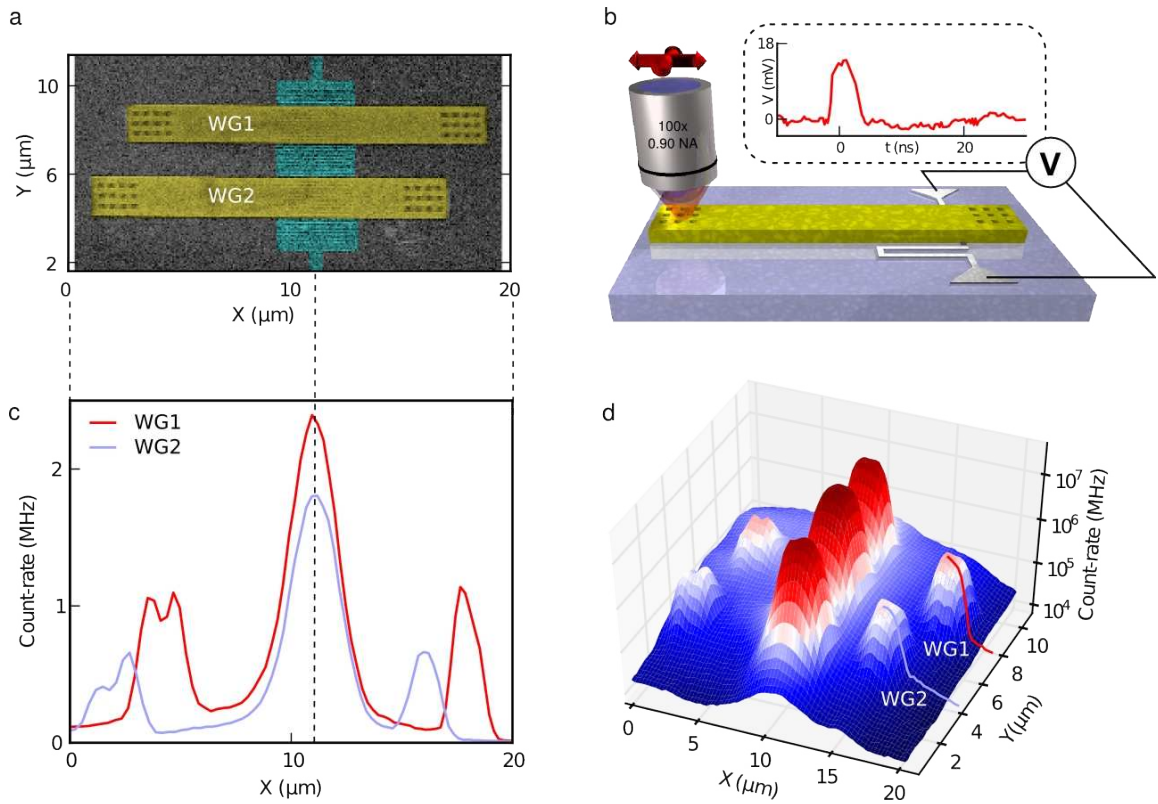

**Fig. 1. a** SEM image showing the superconducting detector (cyan) and two gold waveguides (yellow) with coupling gratings. **b** Representation of the low-temperature setup. The sample is XY-scanned through the laser focus. Illumination of the grating excites plasmons at the substrate/gold interface. After propagating along the waveguide, absorption in the SSPD gives a voltage pulse, *V*. **c** SSPD pulse counts versus laser-spot position. **d** 2D XY-scan. The blue and red lines (WG1 and WG2) indicate where the line-cuts in **c** are taken.

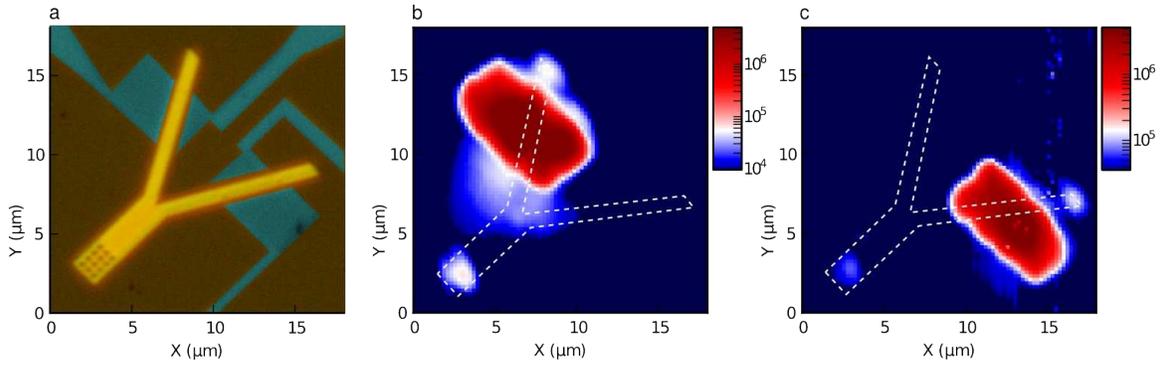

**Fig. 2. a** Microscope image of a plasmon Y-splitter device. The SSPDs (cyan) are colorized. **b, c** Signal of the left (**b**) and right (**c**) detector when scanning a laser (980nm) across the device in **a**. Just one detector produces a signal when illuminating the left or right detector. Both detectors produce a signal when illuminating the grating on the bottom left, indicating that plasmons couple to both arms of the Y-splitter. The white contours of the waveguide are a guide to the eye. Note that the color scales are different due to different dark-count levels in the left and right detectors.

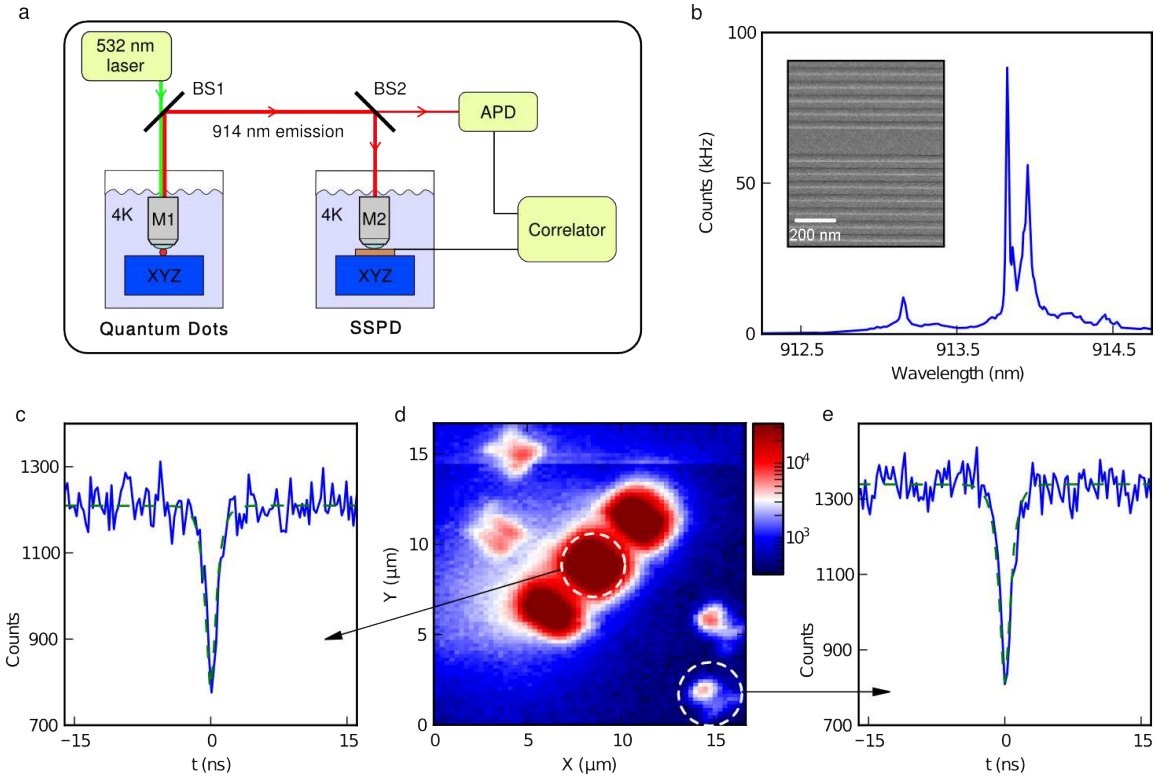

**Fig. 3. a** Schematic representation of the measurement setup. Quantum dots and SSPD are cooled to 4 K in two separate cryostats, which are coupled free-space. BS1 and BS2 are both beamsplitters transmitting 10% and reflecting 90% of the incoming beam. The microscope objectives M1 and M2 are 60x, 0.85 NA and 100x, 0.90 NA respectively. **b** Spectrum of a single SK quantum dot and an SEM image of the distributed Bragg reflector (inset). **c** Time-correlation measurement with 10% of the quantum dot emission going to an APD and 90% sent directly to the SSPD. The integration time is 33 minutes, fitted normalized depth 0.46 and life-time 655 ps. **d** Counts of the SSPD (same device design as Fig. 1a) as a function of focus position of the emission from the quantum dot in **b**. The dashed circles indicate the positions where the correlation measurements in **c** and **e** are taken. **e** Time-correlation measurement with 90% of the light sent to the grating on the bottom right. The dip at t = 0 ns confirms that the photon statistics are maintained after photons are converted into plasmons. The integration time is 8 hours, fitted normalized depth 0.50 and life-time 708 ps.

# Supporting information for "On-Chip Single Plasmon Detection"

Reinier W. Heeres, Sander N. Dorenbos, Benny Koene, Glenn S. Solomon, Leo P. Kouwenhoven, Valery Zwiller

*Methods*

**Sample fabrication**

The SSPDs are fabricated on a sapphire substrate with a thin layer (~5 nm) of NbN. After evaporating Nb/AuPd contacts using an e-beam lift-off step, the meandering detector lines are defined using reactive ion etching with an HSQ e-beam mask. A layer of ~20 nm $SiO_2$ is sputtered to electrically isolate the waveguides from the detectors. A conducting layer is required for the next e-beam step to avoid charging of the isolating sapphire substrate. Therefore a triple layer of photo-resist (Shipley S1813, ~400 nm), tungsten (~7 nm) and PMMA 950k (~100 nm) is used. After writing the waveguide structures, the chip is developed and dry etched ($SF_6$ for tungsten, $O_2$ for photo-resist). Finally, a 125 nm thick layer of gold is evaporated and lifted-off in acetone. A similar fabrication method was reported earlier (see ref. 11 main text).

**Measurement Setup**

The SSPD sample is placed on an Attocube XYZ positioner stage in a dipstick at liquid helium temperature. A Leitz microscope objective (100x, 0.90 NA) is positioned close to the sample and a laser and white-light imaging system is coupled through free-space. The laser source is either a 785 nm or 980 nm diode or a tunable Ti:sapphire laser (Spectra-physics 3900) from 750 to 1000 nm.

The SSPD is current-biased close to the critical current $I_c$ = 12 µA using a home-made bias-T. Pulses are counted using a Stanford Research 400 pulse counter after being amplified by a Miteq JS2-01000200-10-10A low-noise amplifier. Our current devices support count-rates up to 100 MHz. Rates beyond 1 GHz are feasible by reducing the kinetic inductance, using shorter superconducting wires.

After calibrating the forward and backward step sizes of the positioners, the 2D scans are taken by stepping and recording detector counts at each location. For the time-correlation measurements, an amplifier and pulse inverter are added in series, after which the pulses are registered using a Picoquant Picoharp 300 correlator.

**Quantum Dots**

The quantum dot (QD) sample is made by molecular-beam epitaxy and contains a dilute ensemble of self-assembled InAs/GaAs QDs grown at the central anti-node of a planar optical microcavity. The QD density is less than 1 µm$^{-2}$. The microcavity consists of two distributed Bragg reflectors (DBR) surrounding a full-wave GaAs spacer. The DBR region consists of alternating, quarter-wave thick pairs AlAs and GaAs (15.5 lower and 10 upper). The sample was placed in a liquid He bath cryostat with a 4.2K base temperature. At this temperature discrete emission from the QD ensemble spans a wavelength range of approximately 880 to 990 nm. The cavity mode of interest is located at ~915 nm. Discrete single QD emission is coupled to a 60x, 0.85 NA microscope objective (laser spot size ~1 µm) and then to free-space optics. A 10 nm band-pass filter (center wavelength: 920 nm, used at an angle to filter at shorter wavelengths) was used to spectrally select the emission of the QD. The excitation was a standard 532 nm continuous wave diode pumped solid state laser, with approximately 15 µW of power on the sample.

**Fitting Details**

The time correlation data should be fitted with a function of the form $A(1 - \exp(|t - t_0| / \tau))$. However, because the time-response of the system is finite, the exponential is first convolved with a Gaussian $B\exp(-(t / \tau_{TR})^2)$, where B is a normalization constant setting the integral of the Gaussian to 1. The parameter $\tau_{TR}$ is also fitted for, giving the values 305 ps and 345 ps for the fits in Fig. 3c and 3e respectively. This is in good agreement with the expected time-response of the system containing the APD (Perkin Elmer), SSPD and time-correlator (Picoquant Picoharp 300). The APD generates most of the jitter.

## Results

### Other structures

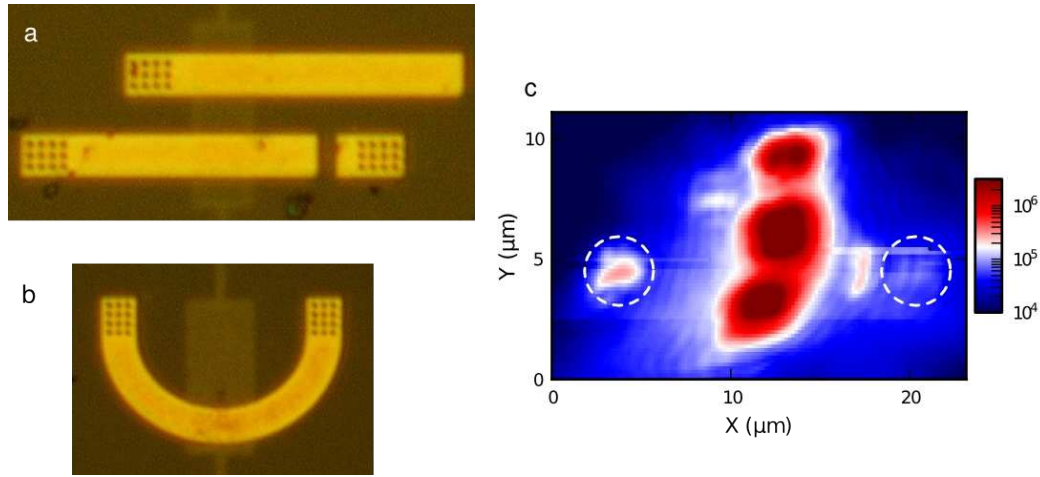

**Fig. S1 a, b** Optical microscope images of two other waveguide structures that have been fabricated, illustrating the flexibility of our approach. **c** XY-scan of the interrupted waveguide in **a**. The measurement shows a large reduction of plasmon detection from the isolated grating on the right (dashed circle, right) compared to the grating on the left (dashed circle, left), in qualitative agreement with FDTD simulations using Lumerical. Note that the 1 µm gap next to the grating on the right is clearly visible in the scan, indicating that it is also an efficient way to excite plasmons.

### Polarization dependence

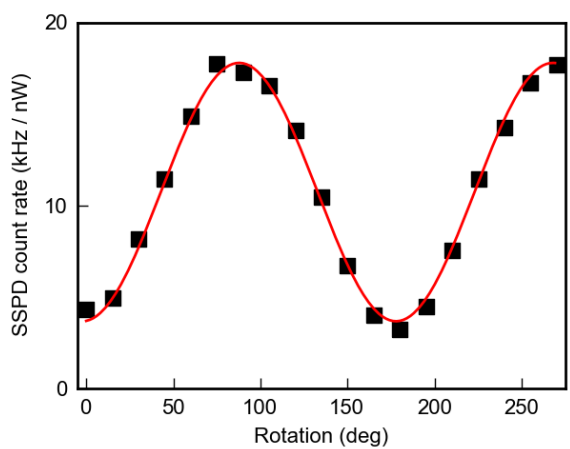

**Fig. S2** SSPD count-rate as a function of the polarization of the laser beam (980 nm), which is varied by rotating a $\lambda/2$ wave-plate. Ninety degrees corresponds to polarization along the waveguide and SSPD meander. The contrast (amplitude / average) is 0.66.

**Power dependence**

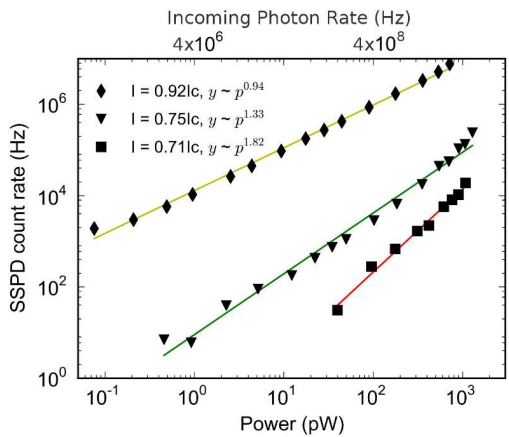

**Fig. S3** Power dependence of the SSPD count-rate as a function of the incident power (785 nm) on the grating closest to the SSPD for different bias currents. $I_c$ = 12 µA is the critical current, at which the detector switches to the normal state. For smaller bias currents, the slope is clearly different from 1. Close to the critical current $I_c$ the fitted line corresponds to an exponent of about 1, indicating a linear dependence. This is a requirement for a process detecting single plasmons. The fact that the curve is linear even at powers where the average number of photons within the dead-time of the detector (~10 ns) is much smaller than one strongly suggests that single plasmons are detected (see ref. 9 main text). The maximum calculated quantum efficiency based on thesecurves is $2.7 \times 10^{-3}$ for the complete photon - plasmon - detection event process.